\documentclass[multphys,vecphys,a4paper]{svmult}

\usepackage{makeidx}         
\usepackage{graphicx}        
\usepackage{multicol}        
\usepackage[bottom]{footmisc}

\makeindex             

\begin{document}

\title*{HI in Local Group analogs:  what does it tell us about 
galaxy formation?}
\titlerunning{HI in Local Group analogs}
\author{D.J. Pisano\inst{1}, D.G. Barnes\inst{2}, B.K. Gibson\inst{3},
L. Staveley-Smith\inst{4}, K.C. Freeman\inst{5}, V.A. Kilborn\inst{6}}

\authorrunning{Pisano et al.}

\institute{Naval Research Laboratory, 4555 Overlook Ave. SW,
Washington, DC 20375 USA; \texttt{pisano@nrl.navy.mil} \and
School of Physics, University of Melbourne, Victoria 3010, Australia; 
\texttt{dbarnes@astro.ph.unimelb.edu.au} \and
Centre for Astrophysics, University of Central Lancashire, Preston, PR1 2HE, 
UK; \texttt{bkgibson@uclan.ac.uk} \and
Australia Telescope National Facility, P.O. Box 76, Epping 
NSW 1710, Australia; \texttt{Lister.Staveley-Smith@atnf.csiro.au} \and
RSAA, Mount Stromlo Observatory, Cotter Road, Weston, ACT 2611, Australia; 
\texttt{kcf@mso.anu.edu.au} \and
Centre for Astrophysics \& Supercomputing, Swinburne University, Hawthorn, 
Victoria 3122, Australia; \texttt{vkilborn@astro.swin.edu.au}}
%
%
\maketitle

\begin{abstract}
We present the results of our HI survey of six loose groups of galaxies 
analogous to the Local Group. The survey was conducted using the Parkes 
telescope and the Australia Telescope Compact Array to produce a census of all 
the gas-rich galaxies and potential analogs to the high-velocity clouds (HVCs) 
within these groups down to $M_{HI} \le$10$^7$M$_\odot$ as a test of 
models of galaxy formation. We present the HI mass function and halo mass 
function for these analogous groups and compare them with the Local Group and 
other environments.  We also demonstrate that our non-detection of HVC analogs 
in these groups implies that they must have low HI masses and be clustered 
tightly around galaxies, including around our own Milky Way, and are not 
distributed throughout the Local Group.
\end{abstract}

\section{Introduction}

As various proceedings from this conference illustrate, the Local Group 
is the nearest and best studied group of galaxies.  We have a census of 
the dwarf galaxies down to very faint magnitude limits.  We can study the
resolved stellar populations  in Local Group galaxies using HST and
large ground-based telescope and re-create their star formation histories.  
These same data can be used to measure distances to these galaxies.  We 
can now even measure the proper motions of some Local Group 
galaxies \cite{brunthaler}.  But just as our position 
within the Milky Way complicates our studies of it, our  position inside the 
Local Group leads to similar problems.  This is particularly true when 
studying the gas within the Local Group that is not associated with stars, 
for example the high-velocity clouds (HVCs)  \cite{WvW}.

The HVCs were first discovered over 40 years ago \cite{muller} and have
remained a mystery since.  They are HI clouds which are moving inconsistent
with simple Galactic rotation with deviation velocities greater than 
90 $km~s^{-1}$.  They also lack stars \cite{willman}, and so we
are generally unable to directly infer their distances.  Without knowing their
distances, we do not know their masses and can not, therefore, divine the
physical processes responsible for their origins.  Most likely, however, they
represent a variety of phenomena.  Some are probably related to a galactic
fountain powered by supernovae explosions \cite{bregman} and are located in 
the lower Galactic halo.  Some HVCs are certainly tidal in origin, like the
Magellanic Stream \cite{putman98}.  Others may be infalling primordial gas,
such as Complex C \cite{wakker}, or gas condensing out of a hot Galactic halo
\cite{maller}.  Finally, some have been suggested to be associated
with the low mass dark matter halos predicted to exist by models of cold dark 
matter (CDM) galaxy formation \cite{blitz,braun}.  

CDM models predict that the Local Group should contain $\sim$300 low mass 
dark halos, while there are only $\sim$20 luminous dwarf galaxies known 
\cite{klypin,moore}.  While this may imply that we lack a complete census of 
the luminous galaxies in the Local Group, it may also be uniquely deficient in 
dwarf galaxies.  Or, perhaps, the HVCs may  populate these dark matter halos 
and solve this ``missing satellite''  problem.

The best way to address these questions is to study other groups of galaxies
in detail.  We can then determine if the population of low mass, gas-rich 
galaxies in the Local Group typical, or if there are relatively more in 
analogous groups?  If HVCs are the solution to the ``missing satellite'' 
problem, then they should be ubiquitous in other groups where we can
then directly determine their masses.  Our HI survey of six loose groups
of galaxies chosen to be analogous to the Local Group attempts to answer
these questions.

\section{Observations}
We selected five of our groups from the optical catalog of \cite{garcia} and
one group from the HIPASS catalog \cite{meyer}.
The resulting groups are between 10.6 and 13.4 Mpc distant, contain
between 3-9 bright, spiral galaxies which are separated, on average, by
$\sim$550 kpc, and have diameters of $\sim$1.6 Mpc; hence the moniker
loose groups.  These groups were known to contain only spiral and irregular 
galaxies and no massive elliptical galaxies; just like the Local Group.  These
groups have no known X-ray emitting gas as is expected to be the case for 
the Local Group.  Their masses, as estimated by the virial theorem, of 
$\sim 10^{11.7-13.6}$M$_\odot$ are comparable to the mass of the Local Group 
$\sim 10^{12.4}$M$_\odot$ \cite{courteau}.

We used the Parkes Multibeam and Australia Telescope Compact Array
(ATCA) to survey the entire area of each group down to a $M_{HI}$
sensitivity of $5-8~\times\,10^{5}~$M$_\odot$ per 3.3 $km~s^{-1}$.  All Parkes
detections in the groups were confirmed to be real by the follow-up
ATCA observations.  A total of 64 HI-rich galaxies were detected in
the six groups, almost twice the number of optically cataloged group
galaxies \cite{garcia} and 50\% more galaxies than were detected by
HIPASS in the same fields \cite{meyer}.  All of our detections
are associated with optical counterparts and  have properties
consistent with typical spiral, irregular, or dwarf irregular
galaxies.  No analogs to the HVCs were detected.

\section{Is the Local Group typical?}

\begin{figure}[ht]
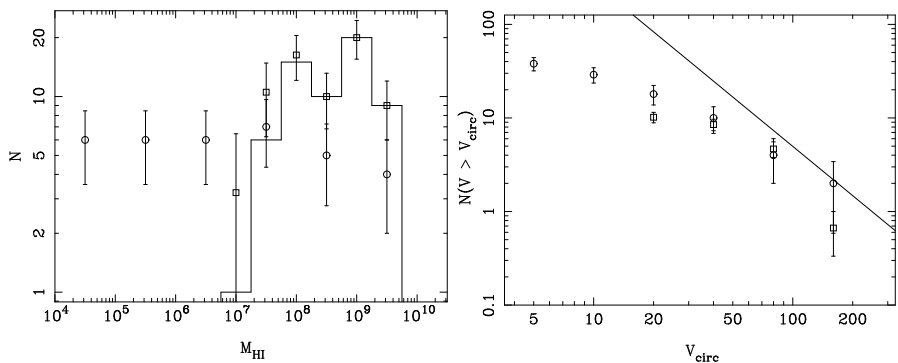

\hfill\includegraphics[angle=-90,width=2.3in]{pisanoF1a.ps}
\hfill\includegraphics[angle=-90,width=2.3in]{pisanoF1b.ps}\hspace*{\fill}
\caption{Left:  The HI mass function (HIMF) of the Local Group
(circles) and the sum of the six loose groups (solid histogram) 
corrected for incompleteness (squares). Right:  The cumulative velocity 
distribution function for the Local Group (circles), and the average 
of the six loose groups (squares).  The solid line represents the CDM model 
from \cite{klypin}, roughly normalized to the second data point.
\label{djpfig:mf}}
\end{figure}

While our sample of loose groups was chosen to be qualitatively analogous
to the Local Group, are the HI properties of these groups quantitatively 
similar?  To answer this, we constructed an HI mass function (HIMF) and
a cumulative circular velocity distribution function (CVDF), both shown
in Figure~\ref{djpfig:mf}.  For the former we have not made any attempt
to quantify the volume observed, so we are simply comparing the slopes.
That is, we are comparing the ratio of high-$M_{HI}$ and low-$M_{HI}$
galaxies in our sample to the Local Group.  The CVDF is a measure of the
dark matter halo mass function as traced by luminous matter, in this case
only for HI-rich galaxies.  As for the HIMF, we are focusing on the slope
of the CVDF and how our groups compare to the Local Group and CDM model 
predictions.  

The HIMF for both the Local Group and our sample of loose groups appears
to be flat; there are the same number of massive and low mass galaxies per
dex of $M_{HI}$ when our sample is corrected for incompleteness down to 
the sensitivity limit of our survey.  This is consistent with other results
for the HIMF in low density environments.  In \cite{zwaan05}, it was shown
that the HIMF flattens in lower density environments with data from HIPASS, 
although others find an opposite trend \cite{springob}.  In \cite{tully}, 
similar results to \cite{zwaan05} were demonstrated from 
a compilation of optical luminosity functions.  This demonstrates that, as far 
as the $M_{HI}$ distribution goes, the Local Group is not atypical.

The CVDF for the Local Group clearly deviates from the solid line representing
the predictions from CDM models \cite{klypin} below 
$V_{circ}~\sim~80~km~s^{-1}$; this is the ``missing-satellite'' problem.  From 
Figure\ref{djpfig:mf}, we can see that our loose groups show the exact same 
deviation and match the CVDF for the Local Group within the statistical errors.
Note that this is the CVDF only for the gas-rich galaxies in these groups, and 
we know that there are more gas-poor systems within the Local Group.  These 
are shown in Figure~\ref{djpfig:mf}, but typically have lower circular 
velocities than the gas-rich systems.  The CVDF for the loose groups has not
been corrected for incompleteness, but we do not expect such corrections to 
be large.  This comparison of CVDFs illustrates, again, that there does not 
appear to a deficit of dwarf galaxies in the Local Group as compared to other 
loose groups.  The Local Group does not appear to be an atypical loose group.  

\section{The nature of the High Velocity Clouds}

Since we can now be confident that our sample of loose groups is truly 
analogous to the Local Group, what does our non-detection of HVCs in these
groups imply for their nature?  To answer this question, we have constructed
a simple model with a Monte Carlo simulation for the distribution of the HVCs 
as described in \cite{pisano}.  For this paper, we use our knowledge of the 
completeness of our survey as a function of integrated flux and linewidth.


Based on this simulation, our non-detection of any HVC analogs in these six 
loose groups implies that they must have relatively low average HI masses, 
$M_{HI}\le~1.6\times~10^5~$M$_\odot$, and be tightly clustered around 
individual galaxies, $D_{HWHM}~\le~100~$ kpc.  These constraints are consistent
with the limits found by others through a variety of other observational 
and theoretical methods \cite{zwaan01,deheij02b,thilker}.  
This also implies that CHVCs are not a major repository of neutral gas, 
$M_{HI}\le~10^8~$M$_\odot$, although there may be a large reservoir of ionized 
gas to fuel star formation in the Milky Way.  These limits are independent of 
the nature of HVCs; they may or may not be associated with dark matter halos
\cite{maller}.  

\section{Conclusions}

We have conducted a deep HI survey of six loose groups of galaxies analogous
to the Local Group.  The survey yielded detections of 64 HI-rich dwarf 
galaxies.  The slopes of the HI mass function and the luminous, dark matter 
halo function for the Local Group and our sample of loose groups are consistent
with each other.  The Local Group, therefore, does not appear to be atypical
in its apparent deficit of low mass, dwarf galaxies as compared to the
predictions of CDM models of galaxy formation.  This does not necessarily 
imply a failure of CDM, but does require, at least, the suppression of 
baryon collapse in low mass halos.  

Our survey also failed to detect any HI clouds lacking stars; no HVC analogs
were detected.  This implies that HVCs are not a large reservoir of neutral 
gas for future star formation in galaxies.  This is consistent with many
models for the origin of HVCs and with other observational limits.  Both 
deeper HI surveys of the $\sim$ 100 kpc environment of individual 
galaxies and higher resolution and sensitivity observations of Milky Way HVCs 
will help to better constrain the nature and origin of HVCs

\bigskip
This research was performed while
D.J.P. held a  National Research Council Research Associateship Award
at the Naval Research  Laboratory.  Basic research in astronomy at the
Naval Research  Laboratory is funded by the Office of Naval Research.
D.J.P. also acknowledges generous support from the ATNF via a Bolton
Fellowship and from NSF MPS Distinguished International Research
Fellowship grant AST 0104439.

%
%

%
%



\printindex


%
%
%
%
%
%
%
%
%
%
%
%
%

\end{document}